\documentclass[aps,prl,twoside,twocolumn,floatfix,superscriptaddress,amsmath,showpacs,amssymb]{revtex4}
\usepackage{amssymb}
\usepackage{bm}
\usepackage{graphicx}
\usepackage{psfrag}
\newcommand{\dar}{\makebox[0mm][l]{\rule{0.33em}{0mm}\rule[0.55ex]{0.044em}{1.55ex}}\rightarrow}

\def\buch{Institute for Nuclear Physics and Engineering, Bucharest, Romania}
\def\buda{KFKI Research Institute for Particle and Nuclear Physics, Budapest, Hungary}
\def\cler{Laboratoire de Physique Corpusculaire, IN2P3/CNRS, and Universit\'{e} Blaise Pascal, Clermont-Ferrand, France}
\def\sp{University of Split, Split, Croatia}
\def\darm{Gesellschaft f\"{u}r Schwerionenforschung, Darmstadt, Germany}
\def\dres{Institut f\"{u}r Strahlenphysik, Forschungszentrum Dresden-Rossendorf, Dresden, Germany} 
\def\heid{Physikalisches Institut der Universit\"{a}t Heidelberg, Heidelberg, Germany}
\def\mosc{Institute for Theoretical and Experimental Physics, Moscow, Russia}
\def\kurc{Kurchatov Institute, Moscow, Russia}
\def\seou{Korea University, Seoul, Korea}
\def\stra{Institut Pluridisciplinaire Hubert Curien and Universit\'{e} Louis Pasteur, Strasbourg, France}
\def\wars{Institute of Experimental Physics, Warsaw University, Warsaw, Poland}
\def\zagr{Ru{d\llap{\raise 1.22ex\hbox{\vrule height 0.09ex width 0.2em}}\rlap{\raise 1.22ex\hbox{\vrule height 0.09ex width 0.06em}}}er Bo\v{s}kovi\'{c} Institute, Zagreb, Croatia}
\def\lan{Institute of Modern Physics, Chinese Academy of Sciences, Lanzhou, China}

\def\mun{Physik Department, Technische Universit\"{a}t M\"{u}nchen, D-85748 Garching, Germany}
\def\jap{Heavy-Ion Nuclear Physics Laboratory, RIKEN, Wako, Saitama 351-0198, Japan}
\def\vien{Institut f\"{u}r Mittelenergie-Physik, \"{O}sterreichische Akademie der Wissenschaften, Boltzmanngasse 3, A-1090, Wien, Austria}

\begin{document}

\title[]{Sub-threshold production of $\Sigma$(1385) baryons \\ in Al+Al collisions at 1.9$A$ GeV}

\author{X.~Lopez} \email{X.Lopez@gsi.de} \affiliation{\darm}
\author{N.~Herrmann} \affiliation{\heid}
\author{P.~Crochet} \affiliation{\cler}
\author{A.~Andronic} \affiliation{\darm}
\author{V.~Barret} \affiliation{\cler}
\author{Z.~Basrak} \affiliation{\zagr}
\author{N.~Bastid} \affiliation{\cler}
\author{M.L.~Benabderrahmane} \affiliation{\heid}
\author{P.~Buehler} \affiliation{\vien}
\author{M.~Cargnelli} \affiliation{\vien}
\author{R.~\v{C}aplar} \affiliation{\zagr}
\author{E.~Cordier} \affiliation{\heid}
\author{P.~Dupieux} \affiliation{\cler}
\author{M.~D\v{z}elalija} \affiliation{\sp}
\author{L.~Fabbietti} \affiliation{\mun}
\author{Z.~Fodor} \affiliation{\buda}
\author{I.~Ga\v{s}pari\'c} \affiliation{\zagr}
\author{Y.~Grishkin} \affiliation{\mosc}
\author{O.N.~Hartmann} \affiliation{\darm}
\author{K.D.~Hildenbrand} \affiliation{\darm}
\author{B.~Hong} \affiliation{\seou}
\author{T.I.~Kang} \affiliation{\seou}
\author{J.~Kecskemeti} \affiliation{\buda}
\author{M.~Kirejczyk} \affiliation{\wars}
\author{Y.J.~Kim} \affiliation{\darm}
\author{M.~Ki\v{s}} \affiliation{\darm} \affiliation{\zagr}
\author{P.~Koczon} \affiliation{\darm}
\author{M.~Korolija} \affiliation{\zagr}
\author{R.~Kotte} \affiliation{\dres}
\author{A.~Lebedev} \affiliation{\mosc}
\author{Y.~Leifels} \affiliation{\darm}
\author{V.~Manko} \affiliation{\kurc}
\author{J.~Marton} \affiliation{\vien}
\author{T.~Matulewicz} \affiliation{\wars}
\author{M.~Merschmeyer} \affiliation{\heid}
\author{W.~Neubert} \affiliation{\dres}
\author{D.~Pelte} \affiliation{\heid}
\author{M.~Petrovici} \affiliation{\buch}
\author{F.~Rami} \affiliation{\stra}
\author{W.~Reisdorf} \affiliation{\darm}
\author{M.S.~Ryu} \affiliation{\seou}
\author{P.~Schmidt} \affiliation{\vien}
\author{A.~Sch\"{u}ttauf} \affiliation{\darm}
\author{Z.~Seres} \affiliation{\buda}
\author{B.~Sikora} \affiliation{\wars}
\author{K.S.~Sim} \affiliation{\seou}
\author{V.~Simion} \affiliation{\buch}
\author{K.~Siwek-Wilczy\'{n}ska} \affiliation{\wars}
\author{V.~Smolyankin} \affiliation{\mosc}
\author{G.~Stoicea} \affiliation{\buch}
\author{K.~Suzuki} \affiliation{\mun}
\author{Z.~Tyminski} \affiliation{\wars}
\author{P.~Wagner} \affiliation{\stra}
\author{E.~Widmann} \affiliation{\vien}
\author{K.~Wi\'{s}niewski} \affiliation{\wars}
\author{D.~Wohlfarth} \affiliation{\dres}
\author{Z.G.~Xiao} \affiliation{\lan}
\author{I.~Yushmanov} \affiliation{\kurc}
\author{X.Y.~Zhang} \affiliation{\lan}
\author{A.~Zhilin} \affiliation{\mosc}
\author{J.~Zmeskal} \affiliation{\vien}

\collaboration{FOPI Collaboration}
\noaffiliation

\date{\today}
\author{P.~Kienle }  \affiliation{\vien} \affiliation{\mun}
\author{T.~Yamazaki} \affiliation{\jap}

\begin{abstract}
First measurement of sub-threshold $\Sigma$(1385) production is presented. Experimental data are presented for Al+Al reactions at 1.9$A$ GeV measured with the FOPI detector at SIS/GSI.
The $\Sigma$(1385)/$\Lambda$ ratio is found to be in good agreement with 
the transport and statistical model predictions.
The results allow for a better understanding of sub-threshold strangeness production and
 strangeness exchange reaction which is the dominant
process for $K^-$ production below and close-to threshold. 
\end{abstract}


\maketitle

Relativistic heavy ion collisions at SIS energies provide interesting
opportunities for studying hot and dense nuclear matter. It allows to 
address fundamental aspects of nuclear physics such as the nuclear-matter 
equation-of-state~\cite{Aichelin:1986ss,Hartnack:2005tr,Sturm:2000dm,Fuchs:2000kp}
and the question whether hadron properties undergo modifications in such an
environment~\cite{chiral}.

Indications for in-medium modifications of charged kaon production and 
propagation
have been experimentally observed by the KaoS~\cite{lastkaos} and the 
FOPI~\cite{Crochet:2000fz,Wisniewski:2001dk} collaborations at SIS energies.
This beam energy range (1 - 2$A$~GeV) is particularly well suited for studying
in-medium properties of strange particles since, as they are produced below
or close-to threshold, their production process is sensitive to 
 nuclear in-medium effects.
While sub-threshold $K^+$ are mostly produced via multi-step processes,
transport models indicate that sub-threshold $K^-$ production results from 
strangeness exchange reactions $\pi + {\rm Y} \leftrightarrow K^- + {\rm B}$
whose rate is intimately linked to the hyperon (${\rm Y} = \Lambda, \Sigma$)
yield~\cite{Fuchs:2005zg,Hartnack:2001zs}.

In addition, recent calculations based on the chiral theory predict an important coupling
of the $K^-$ to the medium via 
$\Sigma(1385)$, $\Lambda(1405)$ and 
$\Lambda(1520)$ resonances~\cite{Cassing:2003vz,Lutz:2001dq,Schaffner-Bielich:1999cp,Tolos:2000fj}
and reveal, in particular, the importance of the $\Sigma(1385)$ hyperon in sub-threshold $K^-$ 
production~\cite{Lutz:2003id}. Theoretical predictions on the in-medium cross-section
 of this strangeness exchange
reaction differ widely~\cite{Cassing:2003vz,Lutz:2001dq,Schaffner-Bielich:1999cp,Tolos:2000fj} and, on the other hand, experimental data on $\Sigma(1385)$
are scarce. \\ 
We report in this letter on the first measurement of charged $\Sigma(1385)$ 
(called $\Sigma^*$ in the following) production in Al+Al collisions at a beam kinetic energy of 
1.9$A$~GeV. 
This measurement corresponds to sub-threshold
production since the threshold beam kinetic energy for $\Sigma^*$ production in 
elementary reaction is $E_{\rm thr.} = 2.33~{\rm GeV}$.
The measured yield is compared to the predictions of statistical
and transport models.
The only other available measurement of $\Sigma^*$ 
production yield in heavy ion collisions has been reported recently 
by the STAR collaboration for the Au+Au system at the top RHIC energy 
($\sqrt{s_{NN}} = 200~{\rm GeV}$) \cite{Adams:2006yu,Salur:2004ar}.\\
The experiment has been performed with the FOPI detector at the SIS Heavy-Ion 
Synchrotron of GSI-Darmstadt by using an Al beam of kinetic energy of 
1.9$A$~GeV on an Al target. The beam intensity was chosen to be 
$8\cdot 10^5$ ions/s and the target thickness was 567~mg/cm$^2$.
The FOPI detector is an azimuthally symmetric apparatus made of several 
sub-detectors which provide charge and/or mass identification over nearly the 
full solid angle. The central part of the detector set-up, placed in a 
super-conducting solenoid, consists of a Central Drift Chamber (CDC) 
surrounded by a barrel of plastic scintillators. The forward part is 
composed of two walls of plastic scintillators 
and a second drift chamber.
For the present analysis, $\Lambda$ and $\Sigma^*$ were reconstructed from
their decay products measured in the CDC ($23^\circ~<~\theta_{lab}~<~114^\circ$).
These are identified by their mass, which is 
determined from the correlation between magnetic rigidity and specific 
energy loss. More details about the configuration and performances of the 
different components of the FOPI apparatus can be found 
in~\cite{Gobbi:1992hw, Ritman:1995td, Andronic:2000cx}.
The events were centrality selected by imposing conditions on the multiplicity
of charged particles detected in the forward part of the detector.
The results presented in the following correspond to the most central 
collisions ($\sigma_{geo}\le315$ mb) representing about 20$\%$ 
of the total geometrical cross section.\\
In order to assess the P($\Sigma^{*-}+\Sigma^{*+}$)/P($\Lambda+\Sigma^{0}$) ratio,
yields of $\Sigma^{*-}$, $\Sigma^{*+}$ as well as primary $\Lambda$ have to be determined.
Note that the reconstructed $\Lambda$ yield includes decay $\Lambda$ from 
$\Sigma^0$ which cannot be isolated with the FOPI detector because $\Sigma^0$ 
decays into a $\Lambda$ and a photon.
While primary $\Lambda$ are reconstructed from the invariant mass of
$(p,\pi^-)$ pair candidates (see~\cite{Ritman:1995tn,mmxl} for more details
about $\Lambda$ reconstruction in FOPI), $\Sigma^*$ are reconstructed
from a topological analysis of their double two body decay:
\begin{eqnarray}
\begin{array}{lllllll}
  \Sigma^{*\pm}~\rightarrow~&\Lambda&\!\!\!\!\!+~\,\pi^\pm & ({\rm BR} = 88\%, c\tau = 5~{\rm fm}) \nonumber \\
  &\dar&p\,\,+~\pi^-  & ({\rm BR} = 64\%, c\tau = 7.89~{\rm cm}).
\end{array}
\end{eqnarray}
Since the short life-time of the $\Sigma^*$ does not allow to disantengle
its decay vertex from the collision vertex, its reconstruction consists
in correlating $\Lambda$ with primary pions.
Therefore, in order to minimize the combinatorial background in the
$(\Lambda,\pi^\pm)$ pair candidates, a high purity sample of $\Lambda$ 
is extracted from the data.
These $\Lambda$ are reconstructed from identified pions and 
protons with the following conditions:

i) a cut on momentum ($p_{lab} \ge 0.1$~GeV/$c$) 
is applied in order to reject particles which are spiraling inside the CDC;

ii) the distance of closest approach (DCA) between tracks and the primary 
vertex in the transverse plane (DCA$_{p}$ $>$ 0.8 cm, DCA$_{\pi^-}$ $>$ 
1.9 cm) is used to enhance the fraction of protons and pions coming from 
a secondary vertex;

iii) a cut on transverse flight distance ($d_{t}<20$~cm) and 
on the pointing angle of the $\Lambda$ ($\Delta\phi<4^{o}$), 
the latter being the difference between the geometrical 
and kinematical azimuthal angles;

iv) a cut on transverse momentum of the $\Lambda$ in the laboratory ($p_{t}\ge0.3$ GeV/$c$)
is used to increase the purity of the signal.  

\begin{figure}[!th]
\vspace{-0.6cm}\includegraphics[width=6.5cm]{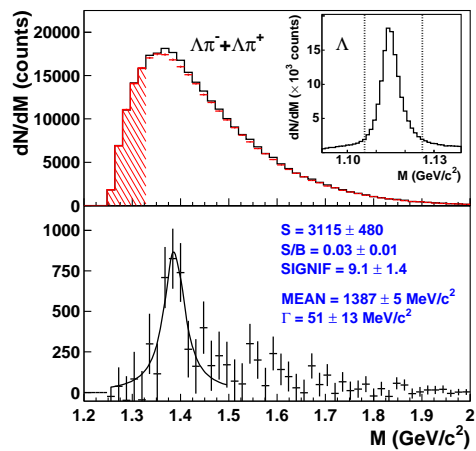}
\caption{\label{fig:minv1} Invariant mass spectra of $p\pi^-$ pairs (inset, 
upper panel) and $\Lambda \pi^{\pm}$ pairs.
The solid histogram and crosses denote the data and the 
scaled mixed-event background, respectively (upper panel). The lower panel
shows the signal after background subtraction. The following characteristics of 
the signal are shown: number of counts in the signal (S), 
signal-to-background ratio (S/B) and significance (SIGNIF). 
The parameters extracted from the fit to the data 
(mean mass value (MEAN) and the width 
($\Gamma$)) are also reported.}
\end{figure}

The invariant mass of the particle pair is calculated from its four-momenta 
at the intersection point of the two tracks. 
The corresponding distribution is shown in the inset of Fig.~\ref{fig:minv1}. 
A total of about 10$^5$ $\Lambda$ are 
reconstructed with a signal-to-background ratio close to 10
from the analysis of 290 million events. 
$\Sigma^{*\pm}$ are then reconstructed by calculating the invariant mass 
of these $\Lambda$ with charged pions after applying a two-sigma mass 
selection cut around the $\Lambda$ nominal mass
(shown by the dashed lines in the inset of Fig.~\ref{fig:minv1}).

The reconstructed $\Sigma^{*-}$ and $\Sigma^{*+}$ are shown in 
Fig.~\ref{fig:minv1}. The combinatorial 
background is obtained with the event-mixing method~\cite{Berger:1976eb}. 
For that purpose, the two decay particles ($\Lambda$ and $\pi^{\pm}$) are 
taken from two different events which present the same particle multiplicity in the CDC. 
In addition, the two events are aligned to the reaction plane in order 
to have the same reference system for both particles. The reaction plane is estimated 
event by event utilizing the transverse momentum procedure 
detailed in~\cite{Danielewicz:1985hn}. 
In the invariant mass range 1.35 to 2 GeV/$c^2$, 16 
excited states of $\Sigma$'s have already been observed~\cite{rev}. 
Therefore, the normalization of the mixed background was 
done in the mass range 1.25 - 1.32~GeV/$c^2$ (grey zone in the upper panel of Fig.~\ref{fig:minv1})
in order not to bias the reproduction of the combinatorial background. 
The shape of the resulting mixed-event background describes the combinatorial 
background and is indicated by crosses in Fig.~\ref{fig:minv1} (upper panel). 
The vertical bars of the crosses correspond to the statistical errors. 
After background subtraction (Fig.~\ref{fig:minv1}, lower panel), the 
remaining peak in the mass spectrum is fitted with a Breit-Wigner function. 
Within the width interval ($\Gamma$), about 3100 $\Sigma^{*\pm}$ 
are found for the applied set of cuts in the analysis. 
The mean mass value 
and width extracted from the fit are in a good agreement, within statistical errors, 
with the values reported by the Particle Data Group~\cite{rev}. 
The width of the $\Sigma^*$ is expected to increase with baryonic density
\cite{oset} but the accuracy of our measurement doesn't permit to draw any conclusion
about a possible broadening. On the right side of the $\Sigma^{*\pm}$ signal, 
some excited states 
are visible as it has been observed in the invariant mass distribution presented 
by the STAR collaboration \cite{Adams:2006yu,Salur:2004ar}. 

 The losses due to decay, acceptance and reconstruction efficiency have to be corrected
in order to extract the particle yields. 
These corrections are determined by means of extensive GEANT simulations 
modeling the full detector response. The IQMD model~\cite{Hartnack:1997ez} is 
used as generator for the underlying heavy-ion event. The $\Sigma^{*\pm}$ and $\Lambda$ 
resonances are embedded in those events (one signal per event) and they are generated separately with 
a momentum distribution according to the Siemens-Rasmussen formula~\cite{sie} 
which describes an expanding system with a temperature $T$ and a radial 
expansion velocity $\beta$. 
The choice of the values of these parameters 
($T=90~{\rm MeV}$, $\beta=0$ and $0.3$) is imposed by previous 
measurements~\cite{Hong:1997mr,mm}. 
Afterwards, a full simulation of the detector, including resolutions in 
energy deposition and spatial position, Front-End-Electronics processing 
and tracking, is performed and the resulting output is subject to the 
same reconstruction procedure as for the experimental data. 
Simulated and experimental spectra of all relevant quantities were
carefully compared.
Since no significant differences were found, we conclude that the apparatus
is properly described by the simulation \cite{xl,mm1}. Finally, the reconstruction 
efficiency is determined by computing the ratio of reconstructed particles 
in the simulation to those initially embedded into the background events.
The systematic error on the $\Lambda$ and $\Sigma^{*\pm}$ yields was
evaluated in two steps.
First the reconstruction efficiency was estimated from simulated
$\Lambda$ and $\Sigma^{*\pm}$ with two values of the radial flow velocity
($\beta=0$ and $\beta=0.3$).
Then the $\Lambda$ and $\Sigma^{*\pm}$ signals were reconstructed under 
different sets of conditions on the relevant quantities previously discussed
($p_{lab}$, DCA, $d_{t}$, $\Delta\phi$ and $p_t$).
In order to minimize the statistical errors, the yields of $\Sigma^{*+}$ and
$\Sigma^{*-}$ are summed and the following ratio can be extracted:
\begin{equation*}
\label{yield}
\frac{\mathrm{P}(\Sigma^{*-}+\Sigma^{*+})}{\mathrm{P}(\Lambda+\Sigma^{0})} = 
0.125\pm 0.026(\mathrm{stat.})\pm 0.033(\mathrm{syst.}).
\end{equation*}
Note that using such a ratio has the additional advantage that the influence 
of the chosen momentum distribution for the simulated signal is cancelled out
to some extent. It is worth to point out that the measured ratio vary, within errors,
by less than a factor two from SIS to RHIC despite the different processes involved
in the $\Sigma^*$ production \cite{Adams:2006yu,Salur:2004ar}.

\begin{figure}[!th]
\vspace{-0.3cm}\includegraphics[width=6.5cm]{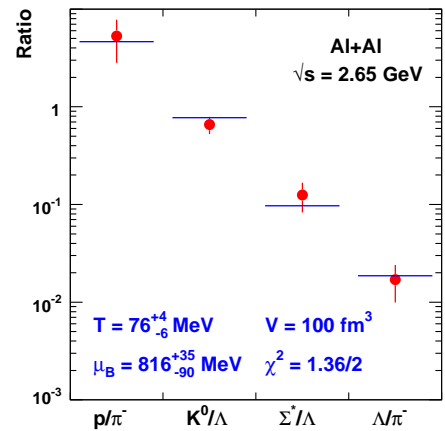}
\caption{\label{fig:comp} Measured hadron yield ratios compared to the thermal model
 calculations for Al+Al collisions at 1.9$A$ GeV. The symbols are data and the lines are
the model calculations.
}
\end{figure}

Our experimental ratio is compared to a statistical model predictions for Al+Al collisions
~\cite{Andronic:2005yp}. The results are shown in Fig.~\ref{fig:comp} where other
particle species such as proton, pion, $\Lambda$ and $K^0$ are included.

The statistical model, based on the canonical ensemble, reproduces the ratios with a temperature of 76 MeV and a baryonic chemical potential of $\mu_{B}=816$ MeV.
This chemical temperature is lower than the kinetical temperature of 95 $\pm$ 10 MeV derived
from previous measurements \cite{Hong:1997mr,mm}. The same model, used to extract thermal parameters from Au+Au collisions at $\sqrt{s_{NN}}=2.7$ GeV~\cite{Andronic:2005yp}, reproduced particle ratios with T = 64 MeV and $\mu_{B}=760$ MeV. It is surprising to notice that at the same energy, a lighter system (Al+Al) brings out higher temperature and baryonic chemical potential, while the volume represents 10\% of the one of the Au+Au system. Finally, a higher temperature appears better suited to describe particle ratios
when including a baryonic resonance and further studies on other strange particles including 
mesonic resonances as $\phi$ and $K$(892) will be useful to constrain the temperature. As it has been reported in \cite{becattini}, an additional parameter as $\gamma_S$ which takes into account a non fully chemical equilibrium of strange particles could also be used to describe all the particle ratios.


The measurement is also compared to the prediction from the UrQMD transport model~\cite{Bleicher:2002rx} for Al+Al collisions \cite{vogel}. It is worth to point out that this model is the only transport code predicting $\Sigma^*$ production at SIS energies. The results of comparisons with statistical and transport models are summarized in Tab.~\ref{tab-1}.

\begin{table}[!h]
\begin{tabular}{|c|c|c|c|}
 \hline
  Yield ratio & Data & Therm. Model & UrQMD  \\
 \hline
  $\frac{{\rm P}(\Sigma^{*-}+\Sigma^{*+})}{{\rm P}(\Lambda+\Sigma^{0})}$ & 0.125 $\pm$ 0.042 &0.097 & 0.177 \\
 \hline
\end{tabular}
\caption{\label{tab-1} Experimental yield ratio and predictions from statistical and
transport models. The error of the experimental result
corresponds to the quadratic sum of statistical and systematic errors.}
\end{table}

The transport model prediction for Al+Al collisions is
in relatively good agreement with the measurement when taking into account the upper limit of the error bars.
The dominant process involved in the UrQMD model
in order to produce $\Sigma^*$ is the fusion of $\Lambda$ and pions (76\%) with an average 
cross section of about 37 mb \cite{vogel}.
The other processes used in the transport code to 
produce $\Sigma^*$ are the interactions between $\Sigma$ and pions (12\%) and, $N^*$ and $\Delta$ 
resonances with baryons (12\%).
 The time dependence of these reactions exhibits a maximum around 8 fm/$c$ which is in agreement 
with the predictions of the IQMD transport model \cite{Hartnack:2001zs} on the time production
 of strange particles such as kaons \cite{hart21}. During the time span between the $\Sigma^*$ production
and the chemical freeze-out, about 8\% of pions coming from the decay of the $\Sigma^*$ are
lost in inelastic interactions. This marginal loss could be
attributed to the small size of the system created in Al+Al collisions.\\

In summary, we have presented new results
about the P($\Sigma^{*-}+\Sigma^{*+}$)/P($\Lambda+\Sigma^{0}$) 
ratio in Al+Al collisions at 1.9$A$~GeV. For the first time, the 
$\Sigma^{*\pm}$ resonances were measured 400~MeV below their production 
threshold.
The experimental result was compared to the predictions of a statistical model
and a transport model.
The measurement is in good agreement with the 
transport model prediction where the dominant reaction to produce the $\Sigma^*$
is the fusion of $\Lambda$ and pions.
On the other hand, the comparison with the thermal model
suggests that a chemical freeze-out temperature of about 80 MeV is needed to describe the 
measured ratio.

The reported measurement should be used as an input for other transport 
codes for testing calculations on sub-threshold strangeness production. In addition, a measurement
of the system size dependence of the $\Sigma^*$ production, as well as other strange resonances, 
will give the opportunity to assess effects of a partial restoration of the chiral symmetry 
at high baryon density that can be reached at SIS energies.\\

We are grateful to M. Bleicher and S.~Vogel for providing us model calculations and for intensive discussions.
 This work was supported by the German BMBF under Contract No.~06HD154, 
by the Korea Science and Engineering Foundation (KOSEF) under grant 
No. F01-2006-000-10035-0, by the mutual agreement between GSI and IN2P3/CEA,
by the Hungarian OTKA under grant No. 47168, within the Framework of the 
WTZ program (Project RUS 02/021), by DAAD (PPP D/03/44611) and by DFG 
(Projekt 446-KOR-113/76/04). We have also received support by the European 
Commission under the 6th Framework Program under the Integrated Infrastructure on:
 Strongly Interacting Matter (Hadron Physics), Contract No.~RII3-CT-2004-506078.

\end{document}